\documentstyle[fleqn,Gc,Epsf]{article}

\begin{document}

\hyphenation{cos-mo-logy}
\hyphenation{asymp-to-tic}
\hyphenation{sti-mu-lating}
\twocolumn[

\Title{
KANTOWSKI-SACHS BRANE COSMOLOGY}

\Author{A.N.\,Makarenko, V.V.\,Obukhov, K.E. Osetrin}
{Lab. for Fundamental Study, Tomsk State Pedagogical University, Russia}

\Abstract{
We consider brane Kantowski-Sachs Universe when bulk space is
five-dimensional Anti-deSitter space. The corresponding
cosmological equations with perfect fluid are written. For several
specific choices of relation between energy and pressure it is
found the behavior of scale factors at early time. In particulary,
for $\gamma=3/2$   Kantowski-Sachs brane cosmology is modified to
become the isotropic one, while for $\gamma=1$ it remains the
anisotropic cosmology in the process of evolution.
}
]
\email 1{andre@tspu.edu.ru}
\email 2{obukhov@tspu.edu.ru}
\email 3{osetrin@tspu.edu.ru}

\section{Introduction}

Since the work by Randall-Sundrum \cite{B1} it was realized
that brane cosmology is quite similar to the standard
four-dimensional cosmology at late times. However, it may significally
differ from standard cosmology at early times. Nevertheless, the
brane gravity shows the newtonian behaviour despite the fact that
braneworld is five-dimensional one.

In the standard as well as in brane cosmology it is quite possible that the early
Universe could be the anisotropic one. During the evolution the anisotropy
stage should quickly be changed by the isotropic stage due to the
classical or quantum matter effects, or due to the
modification of the gravity theory or by some other phenomena. In
the present paper we discuss the brane Kantowski-Sachs (KS) cosmology and
compare it with the standard Kantowski-Sachs cosmology in Einstein gravity
for some specific matter choice. It is shown that brane KS cosmology may
quickly become isotropic one for some choices of matter.

\section{Standard versus brane Kantowski-Sachs cosmology}

2.0. We start from the five-dimensional braneworld which is defined by
the condition $Y(X^I)=0$, where
$I=0,1,2,3,4$ are 5-dimensional coordinates. The starting
action in five dimensional space
is \cite{B2, B3}
\begin{eqnarray}
S&=&\int d^5 X\sqrt{-g_5}\left(\frac{1}{2k_5^2}R_5-\Lambda_5\right)+\nonumber\\
&+&\int_{Y=0}
d^4 x\sqrt{-g}\left(\frac{1}{k_5^2}K^{\pm}-\lambda+L^{matter}\right).
\end{eqnarray}
with $k_5^2=8\pi G_5$ being the 5-dimensional gravitational coupling constant and
$x^\mu$, $(\mu=0,1,2,3)$ are the induced 4-dimensional brane coordinates.
$R_5$ is the 5D intrinsic curvature in the bulk and $K^\pm$ is the intrinsic
curvature on either side of the brane.

The 5D Einstein equation has form
\begin{eqnarray}
^{(5)}G_{IJ}&=&k_5^2 \; \ {}^{(5)}T_{IJ},\nonumber\\
{}^{(5)}T_{IJ}&=&-\Lambda_5 {}^{(5)}g_{IJ}+
\delta(Y)\left[-\lambda g_{IJ}+T_{IJ}^{matter}\right].\nonumber
\end{eqnarray}

Assuming a metric of the form $ds^2=(n_I n_J+g_{IJ})dx^I dx^J$, with
$n_I dx^I=d\xi$ the unit normal to the $\xi=const$ hypersurfaces and $g_{IJ}$
the induced metric on $\xi=const$ hypersurfaces, the effective four-dimensional
gravitational equations on the brane take the form \cite{B2, B3, B4}:
\begin{equation}
\label{qw}
G_{\mu\nu}=-\Lambda g_{\mu\nu}+k_4^2 T_{\mu\nu}+k_5^4 S_{\mu\nu}-E_{\mu\nu},
\end{equation}
where
\begin{eqnarray}
S_{\mu\nu}=\frac{1}{12}T T_{\mu\nu}-\frac{1}{4}T_\mu^\alpha T_{\alpha\nu}
+\frac{1}{24}g_{\mu\nu}(3T^{\alpha\beta}T_{\alpha\beta}-T^2),\nonumber
\end{eqnarray}
and $\Lambda=k_5^2(\Lambda_5+k_5^2\lambda^2/6), \; k_4^2=k_5^2\lambda^2/6$ and
$E_{IJ}=C_{IAJB}n^A n^B$. $C_{IAJB}$ is the 5-dimensional Weyl tensor in the bulk
and $\lambda$ is the vacuum energy on the brane. $T_{\mu\nu}$ is the matter
energy-momentum tensor on the brane with components
$T_0^0=-\rho,\;\;T_1^1=T_2^2=T_3^3=p$ and $T=T_\mu^\mu$ is the trace of the
energy-momentum tensor. One chooses $p=(\gamma-1)\rho$, hence
$1\leq\gamma\leq 2$.

From the equation (\ref{qw}) it follows that there exist several cases:

1) Conventional Einstein Theory (CET) which is 4D and

2) Brane Cosmology (BC) which includes 5D effects.

These cases originate from different relations
between CET and brane world scenario which
can lead to two type of corrections:
(a) the matter fields contribute local "quadratic" energy-momentum correction via
the tensor $S_{\mu\nu}$, and (b) the "nonlocal" effects via bulk Weyl tensor.

We will consider the brane metric in the Kantowski-Sachs form \cite{B5, B6}.
\begin{eqnarray}
ds^2=-dt^2+a_1(t)^2 dr^2+a_2(t)^2(d\theta^2+\sin^2\theta\; d\varphi^2)\nonumber
\end{eqnarray}

The following variables are convenient to introduce:
\begin{eqnarray}
V&=&a_1a_2^2,\nonumber\\
H_i&=&\frac{\dot{a_i}}{a_i},\nonumber\\
H&=&\frac{1}{3}(H_1+2H_2)=\frac{\dot{V}}{3V}.\nonumber
\end{eqnarray}

2.1. In this case the Einstein equation looks:
\begin{equation}
G_{\mu\nu}=-\Lambda g_{\mu\nu}+k_4^2 T_{\mu\nu},\;\;\; \nabla_\mu T^{\mu\nu}=0,\nonumber
\end{equation}
with $G_{\mu\nu}$ the Einstein tensor (4D), $\Lambda$ the cosmological
constant, $k_4$ gravitational coupling $k_4^2=8\pi G$. These equations
for Kantowski-Sachs Universe become:
\begin{eqnarray}
\frac{\dot{a_2}^2}{a_2^2}+2\frac{\dot{a_1}\dot{a_2}}{a_1a_2}+\frac{1}{a_2^2}
&=&\Lambda+k_4^2\rho,\nonumber\\
2\frac{\ddot{a_2}}{a_2}+\frac{\dot{a_2}^2}{a_2^2}+\frac{1}{a_2^2}
&=&\Lambda+k_4^2\rho(1-\gamma),\nonumber\\
\frac{\ddot{a_1}}{a_1}+\frac{\ddot{a_2}}{a_2}+\frac{\dot{a_1}\dot{a_2}}{a_1a_2}
&=&\Lambda+k_4^2\rho(1-\gamma),\nonumber\\
\label{qqq}
\dot{\rho}+\gamma\rho\left(\frac{\dot{a_1}}{a_1}+2\frac{\dot{a_2}}{a_2}\right)
&=&0.
\end{eqnarray}

The Eq. (\ref{qqq}) can be easily solved to describe the time evolution law
of the energy density of the fluid:
\begin{eqnarray}
\rho=\rho_0V^{-\gamma},\;\;\;\rho_0=constant>0.\nonumber
\end{eqnarray}

One can rewrite this equation in another form:
\begin{eqnarray}
\label{q1}
&&\frac{d}{dt}\left(V H_1\right)=\Lambda V+\frac{1}{2}k_4^2\rho_0V^{1-\gamma}(2-\gamma),\\
\label{q2}
&&\frac{d}{dt}\left(V H_2\right)=\Lambda V+\frac{1}{2}k_4^2\rho_0V^{1-\gamma}(2-\gamma)
-a_1,\\
\label{q3}
&&3\dot{H}+H_1^2+2H_2^2=\Lambda+\frac{1}{2}k_4^2\rho_0 V^{-\gamma}(2-3\gamma).
\end{eqnarray}

From the first two equations the equation for V maybe written:
\begin{equation}
\ddot{V}=3\Lambda V+\frac{3}{2}k_4^2\rho_0V^{1-\gamma}(2-\gamma)-2a_1.
\end{equation}

The equation can be partly integrated:
\begin{equation}
\label{qqw}
\dot{V}=\sqrt{3\Lambda V^2+3k_4^2\rho_0V^{2-\gamma}-2\int a_1 dV}.
\end{equation}

From (\ref{q1}) and (\ref{q2}) it follows that:
\begin{eqnarray}
H_1=H+\frac{2}{3V}K,\nonumber\\
H_2=H-\frac{1}{3V}K,\nonumber\\
K=\int a_1 dt.\nonumber
\end{eqnarray}

If we substitute these equations into (\ref{q3}) then:
\begin{equation}
-2a_1 V+\frac{4}{3}\int a_1 dV+\frac{2}{3}\left(\int a_1 dt\right)^2=0.
\end{equation}

Let us consider the asymptotic behaviour. If $V$ is large then solution
has the simple form
\begin{eqnarray}
&&V=a\;e^{\sqrt{3\Lambda}t},\nonumber\\
&&a_1=\frac{a}{b^2}e^{\sqrt{\Lambda/3}t},\,
a_2=b e^{\sqrt{\Lambda/3}t},\\
&&H_1=H_2,\nonumber
\end{eqnarray}
here $a$ and $b$ are constants, $\Lambda$ is not zero.
The behaviour of the system does not depend on $\gamma$ and the anisotropic
CET Universe
becomes isotropic one for large $V$ due to classical matter effects.

The situation for extremely small $V$ is different. The properties of
the CET Universe
will depend on the value of $\gamma$.
From Eq. (\ref{qqw})
\begin{equation}
\dot{V}=\sqrt{3k_4^2\rho_0V^{2-\gamma}+b}.
\end{equation}

For $\gamma=3/2$ the solution takes the form
\begin{eqnarray}
a_1&=&\frac{9a^{2/3}(t-t_0)^{4/9}}{4c^2},\nonumber\\
a_2&=&c(t-t_0)^{4/9},\\
V&=&\frac{9}{4}a^{2/3}(t-t_0)^{4/3}.\nonumber
\end{eqnarray}

Here $c$ is the constant of integration and
$a=\frac{1}{2}k_4^2\rho_0$. For $\gamma=1$ the solution can not be
found in analytical form. Hence, we finished the review of the
anisotropic CET Universe. It is found that even for small
cosmological time the process of isotropisation quickly starts.

2.2. In this subsection we consider the brane cosmology with zero
Weyl tensor which is natural for 5D AdS bulk.
The Einstein equations and evolution law of the energy density take the form
\begin{eqnarray}
\frac{\dot{a_2}^2}{a_2^2}&+&2\frac{\dot{a_1}\dot{a_2}}{a_1a_2}+\frac{1}{a_2^2}=\nonumber\\
&=&\Lambda+k_4^2\rho+\frac{1}{12}k_5^4\rho^2,\nonumber\\
2\frac{\ddot{a_2}}{a_2}&+&\frac{\dot{a_2}^2}{a_2^2}+\frac{1}{a_2^2}=\nonumber\\
&=&\Lambda+k_4^2\rho(1-\gamma)+\frac{1}{12}k_5^4\rho^2(1-2\gamma),\nonumber\\
\frac{\ddot{a_1}}{a_1}&+&\frac{\ddot{a_2}}{a_2}+\frac{\dot{a_1}\dot{a_2}}{a_1a_2}=\nonumber\\
&=&\Lambda+k_4^2\rho(1-\gamma)+\frac{1}{12}k_5^4\rho^2(1-2\gamma),\nonumber\\
\label{qqq1}
&&\dot{\rho}+\gamma\rho\left(\frac{\dot{a_1}}{a_1}+2\frac{\dot{a_2}}{a_2}\right)
=0.
\end{eqnarray}

The Eq. (\ref{qqq1}) can be easily solved:
\begin{eqnarray}
\rho=\rho_0V^{-\gamma},\;\;\;\rho_0=constant>0.\nonumber
\end{eqnarray}

We can rewrite above equations in another form:
\begin{eqnarray}
\label{qq1}
\frac{d}{dt}\left(V H_1\right)&=&\Lambda V
+\frac{1}{2}k_4^2\rho_0V^{1-\gamma}(2-\gamma)+\nonumber\\
&+&\frac{1}{12}k_5^4\rho_0^2V^{1-2\gamma}(1-\gamma),\nonumber\\
\label{qq2}
\frac{d}{dt}\left(V H_2\right)&=&\Lambda V+\frac{1}{2}k_4^2\rho_0V^{1-\gamma}(2-\gamma)+\nonumber\\
&+&\frac{1}{12}k_5^4\rho_0^2V^{1-2\gamma}(1-\gamma)-a_1,\nonumber\\
\label{qq3}
3\dot{H}+H_1^2+2H_2^2&=&\Lambda+\frac{1}{2}k_4^2\rho_0 V^{-\gamma}(2-3\gamma)+\nonumber\\
&+&\frac{1}{12}k_5^4\rho_0^2V^{-2\gamma}(1-3\gamma).\nonumber
\end{eqnarray}

From the first two equations the equation for V can be obtained:
\begin{eqnarray}
\ddot{V}&=&3\Lambda V+\frac{3}{2}k_4^2\rho_0V^{1-\gamma}(2-\gamma)+\nonumber\\
&+&\frac{1}{4}k_5^4\rho_0^2V^{-2\gamma}(1-\gamma)-2a_1.
\end{eqnarray}

It is easy to show that from these equations we can get formally the same
equations that in Conventional Einstein's Theory:

\begin{eqnarray}
&&H_1=H+\frac{2}{3V}K,\nonumber\\
&&H_2=H-\frac{1}{3V}K,\nonumber\\
&&K=\int a_1 dt,\nonumber\\
&&-2a_1 V+\frac{4}{3}\int a_1 dV+\frac{2}{3}\left(\int a_1 dt\right)^2=0.\nonumber
\end{eqnarray}

The asymptotic behaviour for large $V$ has the same form as in CET.
However, for extremely small $V$ one obtains the difference in comparison with CET.

For $\gamma=1$
\begin{eqnarray}
a_1&=&\frac{1}{c_2^2}(c_1+\sqrt{a^2+b}t)^{1/3-\frac{2\sqrt{b}}{3\sqrt{a^2+b}}},\nonumber\\
a_2&=&c_2(c_1+\sqrt{a^2+b}t)^{1/3+\frac{\sqrt{b}}{3\sqrt{a^2+b}}},\nonumber\\
V&=&c_1+\sqrt{a^2+b}t.\nonumber
\end{eqnarray}

Thus, unlike to the situation with CET the analytical solution
appears. Nevertheless, the solution remains anisotropic KS
cosmology.

For $\gamma=3/2$
\begin{eqnarray}
a_1&=&\frac{1}{c_2^2}\left(\frac{3a}{2}\right)^{2/3}(t-t_0)^{2/9},\nonumber\\
a_2&=&c_2(t-t_0)^{2/9},\nonumber\\
V&=&\left(\frac{3a}{2}\right)^{2/3}(t-t_0)^{2/3}.\nonumber
\end{eqnarray}

Here $b,\;t_0,\;c_2$ are constants, $a^2=\frac{1}{4}k_5^4\rho_0^2$.

\section{Discussion}

In summary, we compared the KS brane cosmology with the KS cosmology for Einstein
theory for several classical matter choices.

We demonstrated that with $\gamma=1$ (where CET anisotropic Universe does not have
the analytical asymptotic), brane KS Universe remains anisotropic. For
$\gamma=3/2$, brane KS Universe becomes isotropic at early times as well as
analogous CET case. However, the details of isotropisation (scale factors)
are slightly different, what shows the role of five-dimensional bulk.

One can also consider the simple case, when $a_1=const$. For CET we get the
exact solution
$$a_2=\pm\sqrt{\frac{k_4^2 \rho_0}{a_1^2}-t^2+2t c-c^2}$$
Here $c$ is an integration constant, $\Lambda=0$, $\gamma=2$. However, for
brane cosmology there is no any exact solution. This shows the qualitative
difference between CET and brane cosmology.

It would be really interesting to understand the role of quantum effects
like in Brane New World scenario \cite{B7} to above KS brane cosmology.
In particulary, it is expected that quantum effects may lead to isotropic
cosmology even for initially anysotropic brane Universe as it happens for standard
Einstein theory \cite{B8}.
Another interesting topic could be the generalisation of about discussion for brane 
wormholes (see \cite{B9} for recent discussion).

We thank S.D. Odintsov for very helpful and stimulating discussions.

\end{document}